\documentstyle[psfig]{article}
\begin{document}

\noindent{\Large{\bf The Casimir force for passive mirrors}}

\vspace{0.5cm}

{\sc Astrid Lambrecht$^a$, Marc-Thierry Jaekel$^b$ and Serge Reynaud$^c$} \\

{\it $^a$ Max-Planck-Institut f\"ur Quantenoptik
and Ludwig-Maximilians-Universit\"at}

{\it Hans-Kopfermann-Str.1, D-85748 Garching, Germany}

{\it $^b$ Laboratoire de Physique Th\'{e}orique de l'Ecole Normale
Sup\'{e}rieure \footnote{Unit\'e propre du Centre National de la Recherche Scientifique,
associ\'ee \`a l'Ecole Normale Sup\'{e}rieure et \`a l'Universit\'{e} de Paris
Sud.}}

{\it 24 rue Lhomond, F75231 Paris Cedex 05, France}

{\it $^c$ Laboratoire Kastler Brossel \footnote{Unit\'e de l'Ecole Normale Sup\'{e}rieure et de 
l'Universit\'{e} Pierre et Marie Curie, associ\'ee au Centre National de la Recherche
Scientifique.}}

{\it Universit\'e Pierre et Marie Curie case 74, F-75252 Paris, France}\\

({\sc Physics Letters} {\bf A 225}, p.188 (1997))

\begin{abstract}
We show that the Casimir force between mirrors with arbitrary
frequency dependent reflectivities obeys bounds due to causality and passivity 
properties. The force is always smaller than the Casimir force
between two perfectly reflecting mirrors.
For narrow-band mirrors in particular,
the force is found to decrease with the mirrors bandwidth.\\[3mm]
\end{abstract}

\section{Introduction}

Two reflectors placed in vacuum exert a force onto each other, since the
energy stored between them depends on their relative positions. This well
known Casimir effect is a macroscopic mechanical consequence of quantum
fluctuations of electromagnetic fields. In the standard point of view, the
Casimir energy is deduced from the part of vacuum energy which depends on
the presence and position of reflecting boundaries \cite{Casimir}. In a
local point of view in contrast, the Casimir force is understood as the
radiation pressure exerted upon mirrors by vacuum fluctuations \cite{BM69}.
As it is known from theory and experiments in Cavity Quantum Electrodynamics 
\cite{CQED}, vacuum fluctuations are enhanced or suppressed inside the
cavity, depending on whether the field frequency matches a cavity
resonance or not. Whereas at resonance the vacuum energy density inside the
cavity is increased and thus pushes the mirrors apart, out of resonance
vacuum fluctuations are suppressed and the mirrors are attracted to each
other. The net Casimir force therefore appears as an average between
repulsive and attractive contributions associated respectively with resonant
and antiresonant parts of the spectrum.

In this context, the case of mirrors with frequency dependent reflectivity
amplitudes is particularly interesting, not only because it corresponds to
the realistic case, but also for conceptual reasons. First, it provides a
natural regularisation procedure to dispose of the divergences associated
with the infiniteness of vacuum energy \cite{JP91}. Then, it suggests to
choose specific frequency dependences designed to change the balance between
attractive and repulsive contributions to the force. For narrow-band mirrors
in particular, it may be thought that the Casimir force depends in a
sensitive manner on the relative tuning between the mirrors reflectivity
bands and cavity resonances and, therefore, on the distance between the
mirrors for given reflectivity functions. It might also be hoped that the
force reaches values larger than for perfect mirrors or that it can be
alternatively attractive or repulsive. If confirmed, these features would
allow to design novel and more sensitive experimental demonstrations 
of the Casimir effect \cite{Iacopini93}.

The aim of the present letter is to discuss these expectations and to analyse
them in connection with general properties obeyed by real mirrors, 
namely causality and passivity.
Passivity is a fundamental property related to energy considerations: 
in the absence of an internal gain mechanism,
mirrors are not able to provide energy to the field \cite{Meixner}. 
Passivity of mirrors then ensures stability of intracavity field. In a
first stage we derive properties of the Casimir force evaluated for an
arbitrary passive system of frequency dependent mirrors and show that the
force never exceeds the value corresponding to the limiting case of
perfectly reflecting mirrors. We then concentrate on a cavity formed by two
multilayer dielectric mirrors and find that the Casimir force is always
attractive and a decreasing function of cavity length. We finally consider
narrow-band mirrors designed to disturb the balance between attractive and
repulsive contributions to the Casimir force. We derive a simple expression
for the force revealing it to be in fact much smaller than the value
corresponding to perfect mirrors. In particular the force decreases
with the mirrors bandwidth.

{}For simplicity, we limit ourselves here to calculations in a model
two-dimen\-sio\-nal space-time. As is well known from the analysis of quantum
optical experiments using cavities \cite{Squeez}, each transverse cavity mode
is correctly described by such a model provided that the size of the mirrors
is larger than the spot size associated with the mode and that the mode may
be treated in a paraxial approximation. In the particular case of
narrow-band mirrors, the calculations presented in this paper correspond
therefore to the Casimir force due to a single transverse mode. 
A more elaborate approach would require a detailed evaluation of the effects
of diffraction, accounting in particular for the dependence upon diffraction
of the reflection coefficients of each mode \cite{4D}. The Casimir force
in the realistic four dimensional configuration may be qualitatively
evaluated as the product of the two dimensional result obtained
in this letter with the number of modes efficiently coupled to the cavity, 
that is to say the Fresnel number \cite{Iacopini93}.

\section{The Casimir force for frequency-dependent mirrors}

{}From now on, we restrict our attention to the simple model of a scalar
field in a mono-dimensional space. The Casimir force between
two perfectly reflecting mirrors thus scales as the squared inverse of the
distance $q $ or of the time of flight $\tau $ between the two mirrors \cite
{BM69} 
\begin{equation}
F_{\rm P}=\frac{\pi \hbar }{24c\tau ^2}\qquad \tau =\frac qc  \label{FP}
\end{equation}

More generally, the Casimir force between two partly reflecting and
frequency-dependent mirrors may be expressed as follows \cite{JP91} 
\begin{eqnarray}
F &=&\frac \hbar c\int_0^\infty \frac{{\rm d}\omega }{2\pi }\omega \left(
1-g[\omega ]\right)   \nonumber \\
g[\omega ] &=&\frac{1-\left| r_1[\omega ]r_2[\omega ]\right| ^2}
{\left| 1-r_1[\omega ]r_2[\omega ]e^{2i\omega \tau }\right| ^2}  \nonumber \\
\label{Fg}
\end{eqnarray}
The upper equations describe in a quantitative way the interpretation
presented in the Introduction. They give the Casimir force $F$ as the
difference between outer and inner radiation pressure, defined in such a
manner that a positive value of the force corresponds to attraction. The
density of vacuum energy per mode at frequency $\omega $ is given by the
expressions $\frac{\hbar \omega }{2c}$ and $\frac{\hbar \omega }{2c} g[\omega ]$
respectively outside and inside the cavity. The Airy function $g[\omega ]$
depends on the reflection amplitudes $r_1$ and $r_2$ of the two mirrors and
describes the modification of vacuum energy inside the cavity with respect
to the incoming vacuum energy. It corresponds to an enhancement or
suppression of vacuum fluctuations inside the cavity depending on whether
the field frequency is resonant or non-resonant with a cavity mode \cite
{CQED}. As a result of causality $r_1[\omega ]r_2[\omega]$ is analytic 
for frequencies $\omega$ lying in the upper half $\mathop{\rm Im}\omega >0$ 
of the complex plane \cite{Landau1}. 

The expectations which have been discussed in the Introduction may now be
stated in a more quantitative manner. Since the Airy function $g[\omega ]$
nearly vanishes in the interval between two cavity resonances, the
attractive contribution to the Casimir force from this interval ($\Delta
\omega \simeq \frac \pi \tau $ corresponding to the free spectral range) may
be approximated to the amount $\frac{\hbar \omega }{2c\tau }$. Inside the
cavity resonances in contrast, the Airy function $g[\omega ]$ is much larger
than unity. At exact resonance, it reaches a maximum value proportional to
the cavity finesse. The frequency interval corresponding to the resonance
peaks is much smaller than the free spectral range and the ratio between
these two quantities is controlled by the cavity finesse. As a consequence,
the repulsive contribution corresponding to a resonance peak has nearly the
same magnitude as the attractive contribution corresponding to the
neighbouring free spectral range. This implies that attractive and repulsive
contributions nearly cancel in the evaluation of the net Casimir force.
Actually, for every interval, the attractive and repulsive contributions each
largely exceed the net Casimir force $F_{\rm P}$, at least for high order modes
such that $\omega \tau \gg 1$. This is precisely the reason why it might be
expected that a Casimir force larger than $F_{\rm P}$ could be obtained by
disturbing this fine balance between attractive and repulsive contributions.
We will however show that this expectation cannot be met, because of 
causality and passivity properties of mirrors.

\section{The Casimir force between reciprocal passive mirrors}

To go further in the discussion of the Casimir force, we have now to specify
these properties. In the present section, we will use the minimal assumption 
that the mirrors are causal and passive systems \cite{Meixner}. 
We will first concentrate on the properties of a
single mirror and then deduce the properties of a pair of mirrors and
therefore of the Casimir force.

\begin{figure}[htb]
\centerline{\psfig{figure=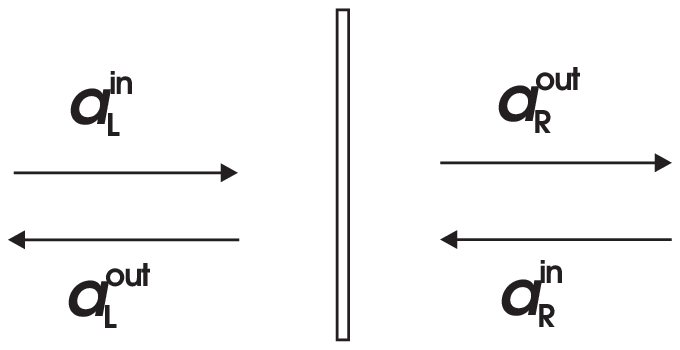,height=3cm}}
\caption{Incoming and outgoing field amplitudes on the left-hand and right-hand side of 
the mirror.}
\label{fig1}
\end{figure}

We consider each mirror as a network with two ports corresponding to the
left-hand and right-hand sides of the mirror. We introduce amplitudes 
$a_{\rm L}^{{\rm in}}$, $a_{\rm L}^{{\rm out}}$, $a_{\rm R}^{{\rm in}}$ and 
$a_{\rm R}^{{\rm out}}$ which
describe the input and output fields at the two ports of the mirror as shown
in Fig.~\ref{fig1}.
These amplitudes are classical numbers representing mean values
of the fields evaluated on the physical boundaries of the mirror. 
The same equation describes also
the transformation of the quantum field amplitudes, that is of the
annihilation and creation operators \cite{JP91}. For dissipative mirrors
extra terms have to be added which describe Langevin forces associated with
loss mechanisms \cite{Kupi92}. We will however not discuss these terms
in more detail since both cases of dissipative or non dissipative mirrors
can be dealt with in exactly the same manner \cite{Barnett96}.

The field amplitudes are gathered in column vectors $a^{{\rm in}}$ and $a^{%
{\rm out}}$ according to the rule 
\begin{equation}
a=\left[ 
\begin{tabular}{l}
$a_{\rm L}$ \\ 
$a_{\rm R}$
\end{tabular}
\right]   \label{defVector}
\end{equation}
We relate input and output field amplitudes through a scattering matrix $S$,
the coefficients of which are two reflection amplitudes $r$ and $\overline{r}$ 
and a transmission amplitude $t$
\begin{equation}
a^{{\rm out}}=Sa^{{\rm in}}  \qquad
S=\left[ 
\begin{tabular}{ll}
$r$ & $t$ \\ 
$t$ & $\overline{r}$%
\end{tabular}
\right]   \label{Smatrix}
\end{equation}
Micro-reversibility of the electromagnetic processes taking place inside the
mirror entails that the network is reciprocal \cite{Onsager}, which
means that the $S$-matrix is symmetrical, the two transmission amplitudes
being equal. However this does not imply that the mirror is symmetrical in the
exchange of its two ports. 
Should this further property be also obeyed, the two reflection amplitudes $r
$ and $\overline{r}$ would be equal too.

We then introduce the magnetic and electric fields at the two ports of the
network (same notations as in eq.(\ref{defVector})) 
\begin{equation}
h=a^{{\rm in}}-a^{{\rm out}}\qquad e=a^{{\rm in}}+a^{{\rm out}}
\end{equation}
These fields are related through an impedance matrix $Z$ 
directly connected to the $S$-matrix 
\begin{equation}
e = Z h  \label{defZ} \qquad
S = \frac{Z-1}{Z+1}  \label{relSZ}
\end{equation}
As an immediate consequence of this relation, it follows that 
\begin{eqnarray}
1-S S^{\dagger} &=& 2\frac{1}{Z+1} \left( Z+Z^{\dagger} \right) \frac{1}{
Z^{\dagger}+1}  \label{relSSZZ}
\end{eqnarray}
We will use this relation in the following to derive an upper
bound of the Casimir force between passive mirrors.

Mirrors are electromagnetic networks which obey properties similar to those
of electrical networks. Passivity of a mirror means
that it does not provide energy to the field when
irradiated by arbitrary incident amplitudes. 
This property may be stated in terms
of the power $\pi$ characterizing this exchange of energy
\begin{eqnarray}
&&\pi (t)=e(t)^{\dagger }h(t)=a^{{\rm in}}(t)^{\dagger }{}a^{{\rm in}}(t)-a^{%
{\rm out}}(t)^{\dagger }{}a^{{\rm out}}(t) \nonumber \\
&&\int_{-\infty }^t\pi (t^{\prime })dt^{\prime }\geq 0 
\end{eqnarray}
The impedance-matrix $Z$ of the mirror, considered as a function of complex
frequency $\omega = ip$, then obeys the following positivity property 
\cite{Meixner} for an arbitrary column vector $h$ 
\begin{eqnarray}
\mathop{\rm Re}p\geq 0\Rightarrow h^{\dagger }\left( Z[ip]+Z[ip]^{\dagger
}\right) h\geq 0  \label{passivity}
\end{eqnarray}
Using relation (\ref{relSSZZ}), the passivity property may equivalently
be written 
\begin{eqnarray}
\mathop{\rm Re}p\geq 0\Rightarrow h^{\dagger }\left( 1-S[ip]S[ip]^{\dagger
}\right) h\geq 0  \label{passivity2}
\end{eqnarray}
By choosing particular vectors $h$ with components 1 and 0 or 0 and 1, one
deduces 
\begin{equation}
\mathop{\rm Re}p\geq 0\Rightarrow \left\{ 
\begin{tabular}{c}
$\left| r[ip]\right| ^2+\left| t[ip]\right| ^2\leq 1$ \\ 
$\left| \overline{r}[ip]\right| ^2+\left| t[ip]\right| ^2\leq 1$%
\end{tabular}
\right\}
\end{equation}
It follows immediately that all scattering coefficients of a passive 
mirror have a modulus smaller than unity for frequencies in the upper
half of the complex plane. Coming back to the notations of equation
(\ref{Fg}), this theorem may be stated 
\begin{eqnarray}
\mathop{\rm Re}p\geq 0\Rightarrow \left| r_j[ip]\right| \leq 1 \label{Th1}
\end{eqnarray}

To evaluate the force (\ref{Fg}), we use the equivalent expression 
\begin{eqnarray}
F &=&\frac \hbar c\int_0^\infty \frac{{\rm d}\omega }{2\pi }\omega \left(
-f[\omega ]-f[\omega ]^{*}\right)  \nonumber \\
f[\omega ] &=&\frac{r_1[\omega ]r_2[\omega ]e^{2i\omega \tau }}
{1-r_1[\omega ]r_2[\omega ]e^{2i\omega
\tau }}=\frac{r_1[\omega ]r_2[\omega ]}{e^{-2i\omega \tau }-
r_1[\omega ]r_2[\omega ]}  \label{F}
\end{eqnarray}
The function $f[\omega ]$ is the loop function corresponding to one
roundtrip of the intracavity field. 
As a consequence of (\ref{Th1}),  the product of the reflection
amplitudes of two passive mirrors has a modulus smaller than unity 
\begin{eqnarray}
\mathop{\rm Re}p\geq 0\Rightarrow \left| r_1[ip]r_2[ip]\right| \leq 1 
\label{Th1a}
\end{eqnarray}
It follows that the system ``cavity (with motionless mirrors) plus vacuum'' 
is stable and $f[\omega ]$ is analytic in
the upper half  of the complex frequency plane. 
The force (\ref{F}) may then be rewritten as an
integral over imaginary frequency $\omega = i p$ with $p$ real 
\cite{Lifshitz56} 
\begin{eqnarray}
F &=&\frac \hbar {\pi c}\int_0^\infty  
\frac{p\ r_1[ip]r_2[ip]}{e^{2p\tau }-r_1[ip]r_2[ip]} {\rm d}p  \label{Fi}
\end{eqnarray}
We have used the fact that $f[ip]$ is real if $p$ is
real since $r_j[ip]$ is the Laplace transform of a real function 
\cite{JP91}. 
In the limit of perfect reflection, expression (\ref{FP}) of
the Casimir force is recovered
\begin{eqnarray}
F_{\rm P} &=&\frac \hbar {\pi c}\int_0^\infty 
\frac{p}{e^{2p\tau }-1} {\rm d}p  \label{FPi}
\end{eqnarray}
With the help of theorem (\ref{Th1a}), we deduce that the
Casimir force (\ref{Fi}) between two arbitrary passive mirrors has an absolute 
value smaller than the value (\ref{FPi}) reached for perfect mirrors 
\begin{eqnarray}
\left| F\right| \leq F_{\rm P} \label{Th1b}
\end{eqnarray}
This is a first bound on the Casimir force which shows that the latter
cannot be arbitrarily modified, because mirrors are passive systems.

\section{The Casimir force between multilayer dielectric mirrors}

We will now concentrate on the particular case of multilayer dielectric
mirrors and derive first the properties obeyed by their reflection
amplitudes from which we will then deduce further properties of the Casimir
force.

We consider the multilayer dielectric mirror to be built of dielectric
slabs. Each slab is symmetrical in the exchange of its two ports. Therefore
the two reflection amplitudes contained in the $S$-matrix of a single slab
are equal. The reflection and transmission amplitudes of a slab denoted by A
are given by the
following relations taken for example from \cite{Siegman} and translated
from real to imaginary frequencies 
\begin{eqnarray}
r_{\rm A}[ip] &=&-\frac{\rho _{\rm A}\left( 1-e^{-2p\xi _{\rm A}}\right) }
{1-\rho_{\rm A}^2e^{-2p\xi _{\rm A}}}  \nonumber \\
t_{\rm A}[ip] &=&\frac{\left( 1-\rho _{\rm A}^2\right) e^{-p\xi _{\rm A}}}
{1-\rho_{\rm A}^2e^{-2p\xi _{\rm A}}}  \nonumber \\
\rho _{\rm A}[ip] &=&\frac{\sqrt{\varepsilon _{\rm A}[ip]}-1}
{\sqrt{\varepsilon _{\rm A}[ip]}+1}  \nonumber \\
\xi _{\rm A}[ip] &=&\sqrt{\varepsilon _{\rm A}[ip]}\frac{l_{\rm A}}c  
\label{slab}
\end{eqnarray}
These amplitudes have the same form as those of a Fabry-Perot cavity where $%
\rho _{\rm A}$ is the reflection amplitude for one interface and $\xi _{\rm A}$ 
an equivalent time of flight between the two interfaces ($l_{\rm A}$ is the 
thickness of the slab). The dielectric constant $\varepsilon _{\rm A}$,
evaluated for imaginary frequencies, approaches unity for $p\rightarrow
\infty $ and is everywhere a real number larger than unity, as a consequence
of Kramers-Kronig relations and passivity of the dielectric medium \cite
{Landau2}.

{}From relations (\ref{slab}) follows that the reflection amplitude of a
dielectric slab is negative for $\mathop{\rm Re}p\geq 0$
while the transmission amplitude is positive.
These latter properties are still fulfilled for multilayer mirrors as we
will show in the following. As every multilayer mirror is built of
dielectric slabs, it remains to proove that the upper properties are
preserved when the slabs are piled up. To this aim we will represent each
mirror by a transfer matrix which relates the fields on its left-hand 
and right-hand side (see Fig.~\ref{fig1})
\begin{equation}
\left[ 
\begin{tabular}{l}
$a_{\rm L}^{{\rm out}}$ \\ 
$a_{\rm L}^{{\rm in}}$
\end{tabular}
\right] =T\left[ 
\begin{tabular}{l}
$a_{\rm R}^{{\rm in}}$ \\ 
$a_{\rm R}^{{\rm out}}$%
\end{tabular}
\right]  \label{defT}
\end{equation}
The explicit form of the $T$-matrix in terms of reflection and transmission 
amplitudes follows directly from expression (\ref{Smatrix}) of the $S$-matrix  
\begin{eqnarray}
T=\frac 1t\left[ 
\begin{tabular}{ll}
$t^2-r\overline{r}$ & $r$ \\ 
$-\overline{r}$ & $1$%
\end{tabular}
\right]  \label{Tmatrix}
\end{eqnarray}
A reciprocal mirror corresponds to a $T$-matrix having a determinant equal
to unity. 

\begin{figure}[htb]
\centerline{\psfig{figure=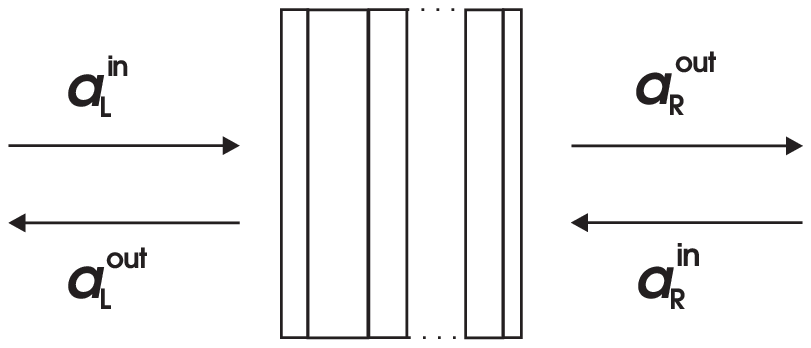,height=3cm}}
\caption{Multilayer mirror obtained by piling up elementary slabs.}
\label{fig2}
\end{figure}

Definition (\ref{defT}) is such that the $T$-matrix associated with a 
multilayer mirror is given by the product of the $T$-matrices
of the elementary slabs which form the mirror (see Fig.~\ref{fig2}).
The $T$-matrix associated with the whole multilayer mirror may thus be obtained 
by iterating piling processes where an elementary slab is added to a given
set of slabs. Denoting the extra slab by A and the set of slabs already piled 
up by B, the iteration step is then described by the relation 
\begin{equation}
T_{\rm AB}=T_{\rm A}T_{\rm B}  \label{chain}
\end{equation}
As a consequence, the reciprocity property is preserved when
slabs are piled up. This characteristic of electromagnetic networks \cite
{Meixner} is only a particular feature of more general properties of
reciprocal networks \cite{Onsager}. In contrast, the symmetry in the
exchange of the two ports which is obeyed by a single slab is no longer
satisfied by multilayer mirrors. Reflection and transmission amplitudes 
are obtained by developing equation (\ref{chain}) with the help of
(\ref{Tmatrix}) 
\begin{eqnarray}
t_{\rm AB} &=&\frac{t_{\rm A}t_{\rm B}}{1-\overline{r}_{\rm A}r_{\rm B}}  
\nonumber \\
r_{\rm AB} &=&r_{\rm A}+\frac{r_{\rm B}t_{\rm A}^2}{1-\overline{r}_{\rm A}
r_{\rm B}}  \nonumber \\
\overline{r}_{\rm AB} &=&\overline{r}_{\rm B}+
\frac{\overline{r}_{\rm A}t_{\rm B}^2}{1-\overline{r}_{\rm A}r_{\rm B}}  
\label{chainAB}
\end{eqnarray}
It immediately follows that the reflection amplitudes evaluated at 
imaginary frequencies for a multilayer dielectric mirror are 
negative while the transmission amplitudes are positive.

Coming back to the notation of equation (\ref{Fg}) we deduce in particular
that the reflection amplitude $r_j$ associated with each mirror is negative 
\begin{eqnarray}
\mathop{\rm Re}p\geq 0\Rightarrow r_j[ip]\leq 0 \label{Th2}
\end{eqnarray}
The product of the reflection amplitudes of
the two mirrors is thus positive 
\begin{eqnarray}
\mathop{\rm Re}p\geq 0\Rightarrow 0\leq r_1[ip]r_2[ip]\leq 1 \label{Th2a}
\end{eqnarray}
It follows that the Casimir force (\ref{Fi}) between two arbitrary 
multilayer dielectric mirrors is attractive for any cavity length
and decreases as a function of the length
\begin{eqnarray}
&&0\leq F\leq F_{\rm P} \nonumber \\
&&\frac{{\rm d}F}{{\rm d}\tau }\leq 0 \label{Th2b}
\end{eqnarray}
These properties constitute further bounds for the Ca\-simir force valid for
any multilayer dielectric mirror.
One may remind at this point that a repulsive Casimir force may be obtained
by considering a cavity built with a dielectric and a magnetic plates \cite
{magnetic}. The product $r$ of the two reflection amplitudes is indeed
negative in this case, so that the force (\ref{Fi}) is repulsive.

\section{The Casimir force between narrow-band mirrors}

One may give a simple and accurate estimation of the Casimir force in the
particular case of narrow-band multilayer dielectric mirrors, i.e. of
mirrors having an optical depth much smaller than the distance between them.

This possibility is based on the existence of a suitable approximation for
the typical variation of the reflection amplitudes as functions of
frequency. For a single slab, the reflection amplitude given in equation (%
\ref{slab}) vanishes at zero frequency $p=0$ and remains small for
frequencies such that $p\xi _{\rm A}<1$. This property has a simple
interpretation. At low frequencies the wavelength of the field is indeed
larger than the optical depth of the slab, so that the slab appears to the
field as a smooth modulation of the refractive index rather than as an
abrupt interface. The reflection amplitude $r_{\rm A}[ip]$ thus vanishes at 
zero frequency
while the transmission amplitude is unity at zero
frequency (cf. eqs (\ref{slab})). The behaviour of the reflection
coefficient at small frequencies is therefore mainly determined by its 
derivative at zero frequency $r_{\rm A}^{\prime }[0]$ 
\begin{eqnarray}
r_{\rm A}[ip] &\simeq &-p\theta _{\rm A}  \nonumber \\
\theta _{\rm A} &=&-ir_{\rm A}^{\prime }[0]  \label{linearapprox}
\end{eqnarray}
We have introduced here a coefficient $\theta _{\rm A}$ which 
can be easily evaluated from (\ref{slab}) for a single slab 
\begin{equation}
\theta _{\rm A}=\frac{\varepsilon _{\rm A}-1}2\frac{l_{\rm A}}c  
\label{thetaslab}
\end{equation}
This coefficient is always positive and is similar to an optical
depth of the dielectric slab, however not measured as a length but as a time.

We can now use transformation rules (\ref{chainAB}) to derive the
corresponding properties of a multilayer dielectric mirror. We first deduce
that its reflection and transmission amplitudes are still respectively zero
and unity. Furthermore it follows from (\ref{chainAB}%
) that the derivatives of the reflection amplitudes at $p=0$ are simply
added when an extra slab is piled up onto a mirror 
\begin{equation}
\theta _{\rm AB}=\theta _{\rm A}+\theta _{\rm B}  \label{thetaA}
\end{equation}
The reflection amplitude of a multilayer dielectric mirror is therefore
still given by the approximated expression (\ref{linearapprox}), with a
coefficient $\theta$ now written as a sum over the multiple layers A
\begin{equation}
\theta =\sum_{\rm A}\frac{\varepsilon _{\rm A}-1}2\frac{l_{\rm A}}c
\label{theta}
\end{equation}
Coefficient $\theta$ is also closely related to the mirrors bandwidth. By using the usual
dispersion relation for the reflection amplitude, we can indeed write
\begin{equation}
\theta =\frac 2\pi \int_0^\infty {\rm d}\omega \frac{\left( -\mathop{\rm
Re}r[\omega ]\right) }{\omega ^2}  \label{bandw}
\end{equation}
In the limit of narrow bandwidth in particular, 
it is directly proportional to 
the bandwidth for mirrors having a given central reflection frequency.

We may now deduce a simple approximation of the Casimir force (\ref{Fi})
between two narrow-band mirrors ($\theta _1\ll \tau $ and $\theta _2\ll \tau $). In the 
integral (\ref{Fi}), 
the dominant contribution indeed comes from low frequencies, due to the
exponential factor $e^{2p\tau }$ appearing in the denominator. This integral
may thus be approximated in terms of the coefficients $\theta _j$ 
\begin{eqnarray}
F &\simeq &\frac \hbar {\pi c}\int_0^\infty p\ r[ip]\ e^{-2p\tau } 
{\rm d}p \nonumber \\
&\simeq &\frac {\hbar \theta _1\theta _2} {\pi c} \int_0^\infty p^3\ 
e^{-2p\tau } {\rm d}p \nonumber \\
&\simeq &\frac{3\hbar \theta _1\theta _2}{8\pi c\tau ^4}  \label{Faa}
\end{eqnarray}
The Casimir force for narrow-band mirrors (\ref{Faa}) can now be compared to
the force evaluated in the limit of perfectly reflecting mirrors (\ref{FP}) 
\begin{equation}
F=F_{\rm P} \frac{9 \theta _1\theta _2}{\pi ^2 \tau ^2}  \label{Fa}
\end{equation}
Clearly the Casimir force between narrow-band mirrors is always much smaller 
than between perfect mirrors. In particular it decreases proportionally to
the product of the reflection bandwidths of the two mirrors.

\section{Discussion}

As discussed in the Introduction, the Casimir force results from a fine
balance between repulsive and attractive contributions associated
respectively with resonant and antiresonant modes of vacuum fluctuations.
This might lead to expect that specific frequency dependences designed to
affect this balance could allow to change quantitative properties of the
Casimir force. We have shown in the present letter that such expectations
cannot be met due to fundamental properties obeyed by real mirrors. For 
any mirror obeying passivity properties, the
Casimir force never exceeds the limiting value obtained for perfect mirrors.
For multilayer dielectric mirrors, even tighter bounds are obtained for the
Casimir force. It remains attractive for any cavity length while its value
is a decreasing function of the cavity length. 

This means that the arguments presented in the Introduction fail despite of
their apparent pertinence. A qualitative interpretation of this failure may
be drawn by recalling that resonances are in fact determined by phase 
conditions. Besides the phase-shift due to free flight between the two
mirrors, the field also undergoes phase-shifts during reflection 
on the mirrors.
Equation (\ref{Fg}) shows that these phase-shifts induce a cavity resonance
shift which was disregarded in the simple discussion
presented in the Introduction. For narrow-band mirrors in particular,
the dephasing of the field differs by a phase $\pi $ above
and below resonance, due to analyticity properties of
the reflection amplitudes.
A phase-shift of $\pi$ is exactly the value required for shifting the
cavity detuning from one mode to the next one. This phase-shift thus forbids
to separate repulsive and attractive contributions to the Casimir
force and, therefore, to meet the expectation of a
more sensitive dependence of the Casimir force versus distance.

These results should be considered in order to assess the feasibility 
of novel experimental demonstrations of the Casimir effect with frequency
dependent mirrors. From the conceptual point of view, they also raise
interesting questions by revealing intimate connections between causality and
passivity properties on one side, and the evaluation of the storage of
vacuum energy by a cavity on the other side.\\ 

\noindent {\bf Acknowledgements}

We are grateful to Paolo Ma\"{\i}a Neto and Paola Puppo
for stimulating discussions.

\end{document}